\documentclass[aps,prc,showpacs]{revtex4}
\begin{document}
\title{On the Eigen Value Problem in Rindler Space}
\author{Sanchita Das$^{a\dagger}$ and Somenath Chakrabarty$^{a,\ddagger}$}
\affiliation{$^a$Department of Physics, Visva-Bharati, Santiniketan 731235,\\
$^\dagger$sanchita.vbphys@gmail,com\\
$^\ddagger$somenath.chakrabarty@visva-bharati.ac.in}
\begin{abstract}
In this article in a very general manner we have investigated the eigen value problem in Rindler space. We have
developed the formalism in an exact form. It has been noticed that although the Hamiltonian is non-hermitian, because of
the $PT$-symmetric nature, the eigen values are real, where $P$ and $T$ are the parity operator and the time
reversal operator respectively. It has further been observed that the
eigen energies are linearly quantized and the binding of the system increases with the increase in the strength of
uniform gravitational field although its origin is purely classical.
\end{abstract}
\maketitle
\section{Introduction}
It is  well known that the conventional Lorentz transformations are
the space-time coordinate transformations between two inertial
frame of references 
\cite{LL}.
However, following the principle of equivalence, it is trivial to obtain the
space-time transformations between a uniformly accelerated frame 
and an inertial frame and vice-versa
in the same manner as it is done in special theory
of relativity
\cite{WB,MTW,RL,MO,BD}. In the present scenario the flat space-time geometry is
called the Rindler space.  
For the sake of  illustration of principle of equivalence,
one may state, that a
reference frame undergoing an accelerated motion in absence of gravity
is equivalent to a frame at rest in presence of 
gravity. Therefore in the present picture, the magnitude of the uniform
acceleration is exactly equal to the strength of constant gravitational field. 
It may be assumed that the gravitational field is produced by a strong gravitating
object. We further approximate that the gravitational field is
constant within a small domain of spacial region. Since it is exactly equal to the uniform
acceleration of the moving frame, this is also called the local acceleration
of the frame.

The article has been organized in the following manner: In the next section, we have reviewed the formalism to obtain
the expression for particle Hamiltonian, when observed from a frame undergoing uniform accelerated motion or in
other words in Rindler space. In section III we have given our formalism on the exact solution of the eigen value
equation and finally in section IV, we have discussed the conclusion of this work.
\section{Special Relativity in Rindler Space}
In this section, for the sake of completeness, 
following the references \cite{MS,MAX,DP} we shall establish some of the useful formulas of
special theory of relativity for a uniformly accelerated frame of
reference. Before we go to the scenario of uniform acceleration of the moving frame, 
let us first assume that the frame $S^\prime$  has rectilinear motion with
uniform velocity $v$ along
$x$-direction with respect to some inertial frame $S$. Further the 
coordinates
of an event occurred at the point $P$ (say) is indicated by
$(x,y,z,t)$ in $S$-frame and with
$(x^\prime,y^\prime,z^\prime,t^\prime)$ in the frame $S^\prime$. The
primed and the un-primed coordinates are related by the conventional form
of Lorentz transformations and are given by
\begin{eqnarray}
x^\prime&=&\gamma(x-vt), ~~y^\prime=y,~~ z^\prime=z ~{\rm{and}}~ \nonumber \\
t^\prime&=&\gamma\left( t-vx\right) ~{\rm{with}}~
\gamma=\left (1-v^2\right )^{-1/2}
\end{eqnarray}
is the well known Lorentz factor. Throughout this article we have followed the natural system of 
units, i.e., speed of light in
vacuum, $c=1$ and later we put the Boltzmann constant $k_B=1$ and the Planck constant $h=1$. 
Next we consider a uniformly accelerated
frame $S^\prime$ moving with uniform acceleration $\alpha$ also along $x$-direction  with respect to $S$-frame. 
Then the Rindler coordinates are given by (see the references \cite{MS,MAX,DP}),
\begin{eqnarray}
t&=&\left (\frac{1}{\alpha}+x^\prime\right )\sinh\left (\alpha t^\prime
\right ) \nonumber  ~~{\rm{and}}~~ \\
x&=&\left (\frac{1}{\alpha}+x^\prime\right )\cosh\left (\alpha t^\prime
\right ) 
\end{eqnarray}
Hence one can also express the inverse relations
\begin{equation}
t^\prime=\frac{1}{2\alpha}\ln\left (\frac{x+t}{x-t}\right )
~~{\rm{and}}~~ x^\prime=(x^2-t^2)^{1/2}-\frac{1}{\alpha}
\end{equation}
The Rindler space-time coordinates as mentioned above
are then just an accelerated frame
transformation of the Minkowski metric of special relativity. The
Rindler coordinate transformations change the Minkowski line element from
\begin{eqnarray}
ds^2&=&dt^2-dx^2-dy^2-dz^2   ~~{\rm{to}}~~ \\ ds^2&=&\left
(1+\alpha x^\prime\right)^2{dt^\prime}^2-{dx^\prime}^2
-{dy^\prime}^2-{dz^\prime}^2
\end{eqnarray}
Since the motion is assumed to be rectilinear and along $x$-direction, 
$dy^\prime=dy$ and $dz^\prime=dz$. The form of the
metric tensor can then be written as
\begin{equation}
g^{\mu\nu}={\rm{diag}}\left (\left (1+\alpha x\right
)^2,-1,-1,-1\right )
\end{equation}
Since we shall deal with the accelerated frame only, we have dropped the prime symbols.
Now following the concept of kinematics of particle motion in special theory
of relativity \cite{LL}, the action
integral may be written as (see also \cite{CGH} and \cite{DLM})
\begin{equation}
S=-\alpha_0 \int_a^b ds\equiv \int_a^b Ldt
\end{equation}
Then using eqns.(5) and (7) and putting $\alpha_0=-m_0$ \cite{LL}, where $m_0$ 
is the
rest mass of the particle, the Lagrangian of the particle is given by
\cite{DLM}
\begin{equation}
L=-m_0\left [\left ( 1+\alpha x\right )^2 -v^2
\right ]
\end{equation}
where $v$ is the velocity of the particle. The momentum of the
particle is then given by
\begin{equation}
p=m_0 v\left [ \left (1+\alpha x \right )^2
-v^2 \right ]^{-1/2}
\end{equation}
Hence the Hamiltonian of the particle or the single particle energy is 
given by
\begin{equation}
H=L-pv=\varepsilon(p)=m_0 \left (1+\alpha x \right ) \left (1+
\frac{p^2}{m_0^2}\right )^{1/2}
\end{equation}
This is the well known Rindler Hamiltonian.
Here $p$ and $v$ are the particle momentum and velocity
respectively along positive $x$-direction. In the inertial frame with
$ \alpha = 0 $, we get back the results of special theory of relativity.
In the case of classical mechanics, $ x$, $p$ and  $H$  are dynamical
variables, whereas in the quantum mechanical picture, $ x$, $p$ and $H $
are operators. In the quantum case $ x$ and  $p $ are also canonical conjugate of each other, i.e., $
[ x , p ] = i \hbar$. Here it is quite obvious that the Hamiltonian operator
represented by eqn.(10) is non-Hermitian. However from our subsequent
analysis and discussion we will show that the eigen values or the eigen
spectra are real in nature. This is found to be solely because of the $PT$ symmetric
nature of the Hamiltonian operator \cite{PT}. 
Under $ P $ and $ T $ operations we have the following relations from $PT$-symmetric 
quantum mechanics: $ P x P^{-1} = -x$, $T x T^{-1} = x$, $P p P ^{-1} = - p$,
$T p T ^{-1} = -p$, $P \alpha P^{-1} = - \alpha$, $T \alpha T^{-1} = \alpha$  and
$T i T ^{-1} = - i$. The last relation is essential for the  preservation
of canonical quantization relation under $PT$ operation, i.e., for
the validity of $ PT [x , p] (PT)^{-1} = i \hbar $.
Hence it is quite obvious to verify that for the Hamiltonian, $ PT H (PT)^{-1} = H $, i.e., the
Hamiltonian operator is $ PT $ invariant. We shall show in our subsequent
discussion that since eigen functions $\Psi$ are the functions of the product $
\alpha x $, which is $ PT $ symmetric, therefore $ PT \Psi (u) = \Psi(u) $,
where $ u $ is a function of the product of $ \alpha x $, i.e., $\psi(u)$ is an eigen function of $PT$ operator with
eigen value $+1$. 
\section{Exact Solution of Eigen Value Problem}
In this section we shall develop a formalism to get exact solution
for the relativistic form of quantum mechanical equation.
In the present formalism we have also used the natural units, i.e., 
$\hbar = c=1$. We begin with the classical Hamiltonian in the 
Rindler space, given by
\begin{equation}
H = (1+\alpha x)(p^2 +m_0^2)^{1/2}
\end{equation}
where $m_0$ is the rest mass of the particle.
Then the eigen value equation $H\Psi=E\Psi$ may be written as 
\begin{equation}
(1 +\alpha x)(-d_x^2 +m_0^2)^{1/2}\Psi = E \psi
\end{equation}
where $ d_x = \frac{d}{dx}$ and we assume that the motion is one 
dimensional and along positive $ x- $direction.
Changing the variable from $ x ~~{\rm{to}}~~ X $, given by
$X = 1 +\alpha x $, 
the above equation reduces to 
\begin{equation}
 X(-d_X^2 +m^{*2})^{1/2} \Psi = E^*\psi
\end{equation}
where we have re-defined  $m^* \rightarrow m_0=m_0/\alpha$ and the
energy eigen value $E^* \rightarrow E=E/\alpha$. 
For the new variable $X$, the limit is from $1$ to $+\infty$, 
instead of $0$ to $+\infty$.
Then we can rewrite the above differential equation as
\begin{equation}
 X(-d_X^2 +m_0 ^2)^{1/2} \Psi = E\psi
\end{equation}
To get an analytical solution, we follow the technique presented in \cite{R10a,R10b,R10c}.
Now using the properties of Dirac delta function, we can write \cite{R10a,R10b,R10c} the left hand side of the
above equation in the form
\begin{equation}
 X (-d_X^2 +m_0^2)^{1/2} \Psi(X) = \int_{-\infty}^{+\infty}
 q(-d_X^2 +m_0^2)^{1/2}\delta(q-X) \Psi(q) dq
\end{equation}
 Since $\delta(x-a)f(x) = \delta(x-a)f(a)$, we have
\begin{equation}
 X (-d_X^2 +m_0^2)^{1/2} \Psi(X)  = \int_{-\infty}^{+\infty}
 q(-d_X^2 +m_0^2)^{\frac{1}{2}}\delta(q-X) \Psi(q) dq
\end{equation}
Using the integral representation of $ \delta -$function, given by
\begin{equation}
\delta(q-X) = \frac{1}{2\pi}\int_{-\infty}^{+\infty} dp \exp[-i(q-X)p]
\end{equation}
where the particle momentum $p$ has been treated here as some kind of integration variable. Then we have
\begin{eqnarray}
 X (-d_X^2 +m_0^2)^{1/2} \Psi(X)  &=& \frac{1}{2\pi}\int_{-\infty}^{+\infty}
\int_{-\infty}^{+\infty}  q(p^2 +m_0^2)^{1/2}\nonumber \\
&&\exp[-i(q-X)p] \Psi(q)
dp dq\nonumber \\
\end{eqnarray}
Hence without the loss of generality, we can re-write the right hand side in the the following form
\begin{eqnarray}
 X (-d_X^2 +m_0^2)^{1/2} \Psi(X) =  
\frac{1}{2\pi}(-d_X^2+m_0^2)&&\int_{-\infty}^{+\infty} q\Psi(q)dq
\nonumber \\
\int_{-\infty}^{+\infty}dp\frac{\exp[-i(q-X)p]}{(p^2 +m_0^2)^{1/2}} 
\end{eqnarray}
with some simple algebraic manipulation, the above equation can be expressed in the following form  \cite{R11}
\begin{eqnarray}
X (-d_X^2 +m_0^2)^{\frac{1}{2}} \Psi(X) & = &
\frac{1}{\pi}(-d_X^2+m_0^2)\int_{-\infty}^{+\infty} q\Psi(q)dq
\int_{0}^{\infty}dp\frac{\cos[(q-X)p]}{(p^2 +m_0^2)^{\frac{1}{2}}}
\nonumber \\ & = & 
\frac{1}{\pi}(-d_X^2 +m_0^2)\int_{-\infty}^{+\infty} dq q \Psi(q)
K_0(m_0|q-X|)
\end{eqnarray}
where $K_0(x)$ is the modified Bessel function of second kind of order zero.
On decomposing the $q$ integral into two parts, we have
\begin{eqnarray}
&&X (-d_X^2 +m_0^2)^{\frac{1}{2}} \Psi(X)  =   
 \frac{1}{\pi}(-d_X^2 +m_0^2)\times \nonumber \\ &&
 \left[\int_{-\infty}^{X} dq q \Psi(q)
K_0[m_0(X-q)] +\int_{X}^{+\infty} dq q \Psi(q) K_0[m_0 (q-X)]\right]
\end{eqnarray}
Then substituting $ X-q = q_1 $ in the first integral and $ q-X = q_2 $ in
the second integral and redefining $ q_1 = q$ in the first integral, 
and $q_2 =q$ in the second integral,
we have 
\begin{eqnarray}
&&X (-d_X^2 +m_0^2)^{\frac{1}{2}} \Psi(X)  =
\frac{1}{\pi}(-d_X^2 +m_0^2)\times \nonumber \\
&&\int_{0}^{\infty} dq
K_0(m_0 q) \left[ (X+q) \Psi(X+q) +(X-q)\Psi(X-q)\right]
\end{eqnarray}
We seek the series solution for the wave function in the form
\begin{equation}
\Psi(X) = \sum_{k=1}^{n+1} \gamma_{k,n+1} X^k \exp(-\beta X)
\end{equation}
 where $\gamma_{k,n+1} ~~{\rm{and}}~~ \beta $ are 
 unknown constants, to be obtained
 from the recursion relations.  
Then the $n_{th}$ term is given by
\begin{eqnarray}
&&X (-d_X^2 +m_0^2)^{1/2} X^n \exp(-\beta X) =
\frac{1}{\pi}(-d_X^2 +m_0^2)\exp(-\beta X)\times \nonumber \\
&&\int_{0}^{\infty} dq
K_0(m_0 q) \left[ (X+q)^{n+1} \exp(-\beta q) +\exp(\beta q)
(X-q)^{n+1}\right] 
\end{eqnarray}
Now expanding $(X+q)^{n+1}$ and $(X-q)^{n+1}$ in Binomial series and then using the standard relation we have
\cite{R11}
\begin{eqnarray}
&&\int_{0}^{\infty} x^{\mu -1} \exp(-\alpha x) K_{\nu}(\beta_1 x
) dx = \nonumber \\
&& \frac{(\pi)^{\frac{1}{2}} (2\beta_1) ^\nu}{(\alpha +\beta_1) ^{\mu
+\nu}}\frac{\Gamma(\mu +\nu)\Gamma(\mu - \nu)}{\Gamma(\mu
+\frac{1}{2})}
F\left(\mu +\nu , \nu +\frac{1}{2}; \mu +\frac{1}{2}; \frac{\alpha -
\beta_1}{\alpha + \beta1}\right) 
\end{eqnarray}
where $F(a,b;c;d) $ is the Hypergeometric function and in our case with $ \nu =0
,~ x=q ,~ \mu -1 = k+1 ,~ \alpha = \beta ~~{\rm{and}}~~ 
\beta_1 =m_0 $, the
integral reduces  to
\begin{equation}
I = \frac{\pi ^{1/2} [\Gamma(k+2)]^2}{\Gamma(k+\frac{5}{2})
(m_0+\beta)^{k+2}}F\left(k+2, \frac{1}{2} ; k+\frac{5}{2} ; -\frac{m_0-
\beta}{m_0 +\beta}\right)
\end{equation}
Then after a little algebra, we have
\begin{eqnarray}
&&{X (-d_X^2 +m_0^2)^{1/2} X^n \exp(-\beta X) = } \nonumber \\
&&\frac{1}{\pi^{1/2}}(-d_X^2 +m_0^2)\exp(-\beta X)\sum_{k=0}^{n+1}
\left(\begin{array} {c} n+1 \\ k \end{array} \right) G_k(m_0,
\beta) X^{n+1-k} 
\end{eqnarray}
where
\begin{eqnarray}
&&G_k(m_0 , \beta) =
\frac{[\Gamma(k+2)]^2}{\Gamma(k+\frac{5}{2})} 
\left[\frac{1}{(m_0+\beta)^{k+2}}F\left(k+2, \frac{1}{2} ; k+\frac{5}{2}
; -\frac{m_0-\beta}{m_0 +\beta}\right)\right ] +\nonumber \\ 
&&\frac{[\Gamma(k+2)]^2}{\Gamma(k+\frac{5}{2})} 
 (-1)^k \left [\frac{1}{(m_0-\beta)^{k+2}}F\left(k+2,\frac{1}{2} ;
k+\frac{5}{2} ; -\frac{m_0+\beta}{m_0 -\beta}\right) \right ]
\end{eqnarray}
Now it is a matter of simple algebra to show that
\begin{eqnarray}
&&X (-d_X^2 +m_0^2) X^{n+1-k} \exp(-\beta X) = \nonumber \\
&&(m_0^2 - \beta ^2)
\exp(-\beta X) X^{n+1-k} + 2(n+1-k) \beta \exp(-\beta X) X^{n-k}
\nonumber \\
&-& (n+1-k)(n-k) \exp(-\beta X) X^{n-k-1}
\end{eqnarray}
Then
\begin{eqnarray}
&&X (-d_X^2 +m_0^2)  X^n \exp(-\beta X) = \nonumber \\
& &\frac{1}{\pi^{1/2}}(-d_X^2 +m_0^2)\exp(-\beta X)\sum_{k=0}^{n+1}
\left(\begin{array} {c} n+1 \\ k \end{array} \right) \nonumber \\
& &\left[(m_0^2 - \beta ^2)G_k(m_0 , \beta) + 2\beta k G_{k-1}(m_0 , \beta)
-k(k-1)G_{k-2}(m_0 , \beta)\right] X^{n+1-k}  \nonumber \\
\end{eqnarray}
Hence from the quantum mechanical equation
\begin{equation}
X (-d_X^2 +m_0^2)^{1/2} \Psi (X) = E\Psi(X) ,
\end{equation}
we have from the series solution (polynomial form) of $ \Psi(X) $,
\begin{eqnarray}
\lefteqn{X(-d_X^2 +m_0^2)^{1/2} \sum_{k=1}^{n+1} \gamma_{k,n+1} X^k \exp(-\beta
X) }\nonumber \\
& = &\sum_{k=1}^{n+1} \gamma_{k,n+1} \sum_{p=0}^{k} F_{p,k}(m_0,\beta)
X^{k+1-p}  \nonumber \\
& = & E_n \sum_{k=1}^{n+1} \gamma_{k,n+1}\exp(-\beta X) X^k
\end{eqnarray}
The expressions for $F_{p,k} (m_0,\beta)$ 
and $\gamma_{k,n+1}$  can be defined in terms of the parameter $\beta$ and the rest mass $m_0$ \cite{R10a}.
From the above equation, equating the coefficient of $x^2$, we have
\begin{equation}
E_n = \frac{\gamma_{1,n+1}}{\gamma_{2,n+1}} F_{0,n+1}(m_0 ,\beta)
\end{equation}
the energy corresponding to the $n_{th}$ level of the spectrum. 
The quantity $\gamma_{l,n}$ can be obtained from \cite{R10a}.
Further equating the coefficients of $X^l$ from both the sides and
putting $p=1$, we have
\begin{eqnarray}
&&\gamma_{l,n+1}(\beta,m_0) F_{1,l}(m_0 , \beta) = E_l(\beta,m_0) \gamma_{l,n+1}(m_0,\beta) ~~{\rm{or}} \\
&& E_l(m_0,\beta) = F_{1,l} (m_0 ,\beta)
\end{eqnarray}
which gives the energy spectrum in terms of the unknown 
parameter $\beta$, quantum number $l$ and the strength of gravitational
field $\alpha$.  Where
\begin{equation}
F_{1,l}(m_0,\beta)=\frac{-l\beta \alpha}{(2\pi m_0)^{1/2}(m_0 ^2 - \beta
^2)^{3/4}} P_{-\frac{1}{2}}^{-\frac{3}{2}}\left
(\frac{\beta}{m_0}\right ) \\
\end{equation}
where $l=1,2,3,......$, positive integers.
Therefore to obtain the energy spectrum, we have to evaluate the
associated Legendre function. It is to be noted further that the energy
spectrum is real and linearly quantized ($\propto l$). 
Now because of the factor $\exp(-\beta X)$ the wave functions are bounded
and also the energy eigen values are negative in nature. 
From the above
expression it is quite obvious that $\beta < m_0$. This is also a
necessary condition for the argument $z$ of $P_\nu^\mu(z)$ to be
real.
It has been noticed that the magnitude of the energy eigen 
value $\mid E_l(m_0,\beta)\mid $ increases with $\beta$ and the rise is
very sharp as $\beta \longrightarrow m_0$. However, it is actually 
negative throughout and the negativity increases with $\beta$ and also with
$\alpha$. Further,
the eigen functions are $\propto \exp(-\beta X)$, therefore with the increase of $\beta$ the wave function converges
to zero very quickly, whereas the eigen states become more bound
because of high negative value of energy for large $\beta$. 
Or in other wards, when $\beta/m_0$ is very close to unity, the nature of
wave functions and energy eigen values indicate that the binding of the states are strong enough.
Therefore with the increase of negative value of the energy makes the state more bound and simultaneously the
spread of wave function in space is reduced. The later is also in 
agreement with more stronger binding. At this point we would like to 
emphasize that with the increase in the strength of gravitational 
field produced by a black, i.e., as we approach more and more close to
 the event horizon, the absolute value of the energy eigen value 
increases more and more makes the system more bound. In our study since
$\alpha$ is the gravitational field strength of the black hole, we may 
conclude that the increase in classical background gravitational field, 
which is assumed to be uniform locally, makes the particle, which is 
quantum mechanical in nature, more strongly bound. The binding is with 
the black hole in presence of strong classical background field. 

\end{document}